\documentclass[11pt,a4paper]{article}
\usepackage{amsfonts}
\usepackage{amsmath,amssymb}
\usepackage{epsfig,graphicx}

\begin{document}

\begin{flushright}
CERN-PH-TH/2007-188
\end{flushright}

\begin{center}
{\LARGE Constrained Gauge Fields from Spontaneous Lorentz Violation }

\bigskip

\textbf{J.L.~Chkareuli}$^{1}$, \textbf{C.D.~Froggatt}$^{2}$, \ \textbf{J.G.
Jejelava}$^{1}$\textbf{\ and H.B. Nielsen}$^{3,4}$

\bigskip

$^{1}$\textit{E. Andronikashvili} \textit{Institute of Physics and }

\textit{I. Chavchavadze State University, 0177 Tbilisi, Georgia\ \vspace{0pt}%
\\[0pt]
}

$^{2}$\textit{Department of Physics and Astronomy, Glasgow University,}\\[0pt%
]
\textit{Glasgow G12 8QQ, Scotland}\vspace{0pt}

$^{3}$\textit{Niels Bohr Institute, Blegdamsvej 17-21, DK 2100 Copenhagen,
Denmark}\\[0pt]

$^{4}$\textit{TH Division, Physics Department, CERN, Geneva, Switzerland}\\[%
0pt]
\bigskip \bigskip \bigskip \bigskip \bigskip \bigskip \bigskip

\textbf{Abstract}

\bigskip
\end{center}

Spontaneous Lorentz violation realized through a nonlinear vector field
constraint of the type $A_{\mu }A^{\mu}=M^{2}$ ($M$ is the proposed scale
for Lorentz violation) is shown to generate massless vector Goldstone
bosons, gauging the starting global internal symmetries in arbitrary
relativistically invariant theories. The gauge invariance appears in essence
as a necessary condition for these bosons not to be superfluously restricted
in degrees of freedom, apart from the constraint due to which the true
vacuum in a theory is chosen by the Lorentz violation. In the Abelian
symmetry case the only possible theory proves to be QED with a massless
vector Goldstone boson naturally associated with the photon, while the
non-Abelian symmetry case results in a conventional Yang-Mills theory. These
theories, both Abelian and non-Abelian, look essentially nonlinear and
contain particular Lorentz (and $CPT$) violating couplings when expressed in
terms of the pure Goldstone vector modes. However, they do not lead to
physical Lorentz violation due to the simultaneously generated gauge
invariance.

\thispagestyle{empty}\newpage

\section{Introduction}

One of the most interesting examples where quantum field theory might
provide some guiding rules for the search for new physics could be that of
the origin of internal symmetry patterns in particle physics owing to
space-time properties at very small distances. In this connection, the
relativistic or Lorentz invariance seems to play a special role with respect
to the observed internal local symmetries. The old idea \cite{bjorken} that
spontaneous Lorentz invariance violation (SLIV) may lead to an alternative
theory of QED, with the photon as a massless vector Nambu-Goldstone boson,
still remains extremely attractive in numerous theoretical contexts \cite%
{book} (for some later developments, see the papers \cite{cfn}). At the same
time, Lorentz violation on its own has attracted considerable attention in
recent years as an interesting phenomenological possibility appearing in
various quantum field and string theories [4-9]. Actually, the SLIV idea is
in accordance with superstring theory, particularly with the observation
that the relativistic invariance could spontaneously be violated in
superstrings \cite{alan1}.

The first models realizing the SLIV conjecture were based on the four
fermion (current-current) interaction, where the gauge field appears as a
fermion-antifermion pair composite state \cite{bjorken}, in complete analogy
with the massless composite scalar field in the original Nambu-Jona-Lazinio
model \cite{NJL}. Unfortunately, owing to the lack of a starting gauge
invariance in such models and the composite nature of the Goldstone modes
which appear, it is hard to explicitly demonstrate that these modes really
form together a massless vector boson as a gauge field candidate. Actually,
one must make a precise tuning of parameters, including a cancellation
between terms of different orders in the $1/N$ expansion (where $N$ is the
number of fermion species involved), in order to achieve the massless photon
case (see, for example, the last paper in \cite{bjorken}). Rather, there are
in general three separate massless Goldstone modes, two of which may mimic
the transverse photon polarizations, while the third one must be
appropriately suppressed.

In this connection, a more instructive laboratory for SLIV consideration
proves to be a simple class of QED type models [11-14] having from the
outset a gauge invariant form. In these models the spontaneous Lorentz
violation is realized through the nonlinear dynamical constraint $A_{\mu
}A^{\mu }=n_{\nu }n^{\nu }M^{2}$ (where $n_{\nu }$ is a properly oriented
unit Lorentz vector, $n_{\nu }n^{\nu }=\pm 1$, while $M$ is the proposed
SLIV scale) imposed on the starting vector field $A_{\mu }$, in much the
same way as it occurs for the corresponding scalar field in the nonlinear $%
\sigma $-model for pions \cite{GL}. Note that a correspondence with the
nonlinear $\sigma $-model for pions may be somewhat suggestive, in view of
the fact that pions are the only presently known Goldstones and their
theory, chiral dynamics \cite{GL}, is given by the nonlinearly realized
chiral $SU(2)\times SU(2)$ symmetry rather than by an ordinary linear $%
\sigma $-model. The above constraint means in essence that the vector field $%
A_{\mu }$ develops some constant background value $<A_{\mu }(x)>$ $=n_{\mu
}M $\ and the Lorentz symmetry $SO(1,3)$ formally breaks down to $SO(3)$ or $%
SO(1,2)$ depending on the time-like ($n_{\nu }n^{\nu }>0$) or space-like ($%
n_{\nu }n^{\nu }<0$) nature of SLIV. This allows one to explicitly
demonstrate that gauge theories, both Abelian and non-Abelian, can be
interpreted as spontaneously broken theories[11-14], although the physical
Lorentz invariance still remains intact.

However, the question naturally arises of whether a gauge symmetry is
necessary to start with. If so, this would in some sense depreciate the
latter approach as compared with those of the original composite models \cite%
{bjorken}, where a gauge symmetry was hoped to be derived (while this has not
yet been achieved). Remarkably, as we will see, it happens that one does not
need to specially postulate the starting gauge invariance, when considering
the nonlinear $\sigma $-model type spontaneous Lorentz violation in the
framework of an arbitrary relativistically invariant Lagrangian for
elementary vector and matter fields, which are proposed only to possess some
global internal symmetry. In the present article we start by a priori only
assuming a global symmetry but no gauge invariance, taking all the terms in
the Lagrangian allowed by Lorentz invariance. With such a Lagrangian, the
vector field $A_{\mu }$ typically develops a non-zero vacuum expectation
value, 
\begin{equation}
<A_{\mu }(x)>=n_{\mu }M.  \label{Avev}
\end{equation}%
In the limit analogous to the approximation of the linear $\sigma $-model by
the nonlinear $\sigma $-model, we get the nonlinear constraint\footnote{%
Actually, some way to appreciate a possible origin for the supplementary
condition\ (\ref{con}) might be by the inclusion of a \textquotedblleft
standard\textquotedblright\ quartic vector field potential $U(A_{\mu })=-%
\frac{m_{A}^{2}}{2}A^{2}+\frac{\lambda _{A}}{4}(A^{2})^{2}$ in the vector
field Lagrangian, as can be motivated to some extent \cite{alan1} from
superstring theory. This potential inevitably causes the spontaneous
violation of Lorentz symmetry in a conventional way, much as an internal
symmetry violation is caused in a linear $\sigma $ model for pions \cite{GL}%
. As a result, one has a massive \textquotedblleft Higgs" mode (with mass $%
\sqrt{2}m_{A}$) together with massless Goldstone modes associated with the
photon. Furthermore, just as in the pion model, one can go from the linear
model for the SLIV to the non-linear one by taking the limit 
$\lambda_{A}\rightarrow \infty ,$ $m_{A}^{2}\rightarrow \infty $ (while keeping the
ratio $m_{A}^{2}/\lambda _{A}$ to be finite). This immediately leads to the
constraint (\ref{con}) for the vector potential $A_{\mu }$ with $%
n^{2}M^{2}=m_{A}^{2}/\lambda _{A}$, as appears from the validity of its
equation of motion. Another motivation for the nonlinear vector field
constraint (\ref{con}) might be an attempt to avoid an infinite self-energy
for the electron in classical electrodynamics, as was originally suggested
by Dirac \cite{dir} and extended later to various vector field theory cases 
\cite{vent}.} 
\begin{equation}
A^{2}=n^{2}M^{2}\qquad (A^{2}\equiv A_{\mu }A^{\mu },\quad n^{2}\equiv
n_{\nu }n^{\nu }).  \label{con}
\end{equation}%
In this paper we shall simply postulate that the existence of the constraint 
(\ref{con}) is to be upheld by adjusting the parameters of the Lagrangian. We
then show that the SLIV conjecture, which is related to the condensation of
a generic vector field or vector field multiplet, happens by itself to be
powerful enough to impose gauge invariance, provided that we allow the
corresponding Lagrangian density to be adjusted to ensure self-consistency
without losing too many degrees of freedom. Due to the Lorentz violation,
this theory acquires on its own a gauge-type invariance, which gauges the
starting global symmetry of the interacting vector and matter fields
involved. In essence, the gauge invariance (with a proper gauge-fixing term)
appears as a necessary condition for these vector fields not to be
superfluously restricted in degrees of freedom. In fact the crucial
equations (\ref{id}) and (\ref{id1}) below express the relations needed to
reduce the number of independent equations among the equations of motion and
the constraint (\ref{con}). But notice that we are not assuming gauge
invariance to derive equations (\ref{id}) and (\ref{id1}); our philosophy is
to derive gauge invariance not to put it in. Due to the constraint (\ref{con}%
), the true vacuum in a theory is chosen by the Lorentz violation, SLIV. The
self-consistency problem to which we adjusted the couplings in the
Lagrangian might have been avoided by using a Lagrange multiplier associated
with the constraint (\ref{con}). However it is rather the philosophy of the
present article to look for consistency of the equations of motion and the
constraint, without introducing such a Lagrange multiplier.

In the next Sec.~2 we consider the global Abelian symmetry case, which
eventually appears as ordinary QED taken in a nonlinear gauge. While such a
model for QED was considered before on its own [11-14], we actually derive
it now using the pure SLIV conjecture. Then in Sec.~3 we generalize our
consideration to the global non-Abelian internal symmetry case and come to a
conventional Yang-Mills theory with that symmetry automatically gauged.
Specifically, we will see that in a theory with a symmetry group $G$ having $%
D$ generators not only the pure Lorentz symmetry $SO(1,3)$, but the larger
accidental symmetry $SO(D,3D)$ of the Lorentz violating\ vector field
constraint also happens to be spontaneously broken. As a result, although
the pure \ Lorentz violation still generates only one true Goldstone vector
boson, the accompanying pseudo-Goldstone vector bosons related to the $%
SO(D,3D)$ breaking also come into play properly completing the whole gauge
field multiplet of the internal symmetry group taken. Remarkably, they
appear to be strictly massless as well, being protected by the
simultaneously generated non-Abelian gauge invariance. When expressed in
terms of the pure Goldstone vector modes these theories, both Abelian and
non-Abelian, look essentially nonlinear and contain Lorentz and $CPT$
violating couplings. However, due to cancellations, they appear to be
physically indistinguishable from the conventional QED and Yang-Mills
theories. On the other hand, their generic, SLIV induced, gauge invariance
could of course be broken by some high-order operators, stemming from very
short gravity-influenced distances that would lead to the physical Lorentz
violation. This and some other of our conclusions are discussed in the final
Sec.~4.

\section{\protect\bigskip Abelian theory}

Suppose first that there is only one vector field $A_{\mu }$ and one complex
matter field $\psi $, a charged fermion or scalar, in a theory given by a
general Lorentz invariant Lagrangian $L(A,\psi )$ with the corresponding
global $U(1)$ charge symmetry imposed. Before proceeding further, note first
that, while a conventional variation principle requires the equation of
motion 
\begin{equation}
\frac{\partial L}{\partial A_{\mu }}-\partial _{\nu }\frac{\partial L}{%
\partial (\partial _{\nu }A_{\mu })}=0  \label{eqm}
\end{equation}%
to be satisfied, the vector field $A_{\mu }$, both massive and massless,
still contains one superfluous component which is usually eliminated by
imposing some supplementary condition. This is typically imposed by taking
the 4-divergence of the Euler equation (\ref{eqm}). Such a condition for the
massive QED case (with the gauge invariant $F_{\mu \nu }F^{\mu \nu }$ form
for the vector field kinetic term) is known to be the spin-1 or Lorentz
condition $\partial _{\mu }A^{\mu }=0$, while for the conventional massless
QED many other conditions (gauges) may alternatively be taken.

Let us now subject the vector field $A_{\mu }(x)$ in a general Lagrangian $%
L(A_{\mu},\psi )$ to the SLIV constraint (\ref{con}), which presumably
chooses the true vacuum in a theory. Once the SLIV constraint is imposed,
any extra supplementary condition is no longer possible, since this would
superfluously restrict the number of degrees of freedom for the vector field
which is inadmissible. In fact a further reduction in the number of
independent $A_{\mu }$ components would make it impossible to set the
required initial conditions in the appropriate Cauchy problem and, in
quantum theory, to choose self-consistent equal-time commutation relations%
\footnote{%
For example the need for more than two degrees of freedom is well-known for
a massive vector field and for quantum electrodynamics. In the massive
vector field case there are three physical spin-1 states to be described by
the $A_{\mu }$, whereas for QED, apart from the two physical (transverse)
photon spin states, one formally needs one more component in the $A_{\mu }$ (%
$A_{0}$ or $A_{3}$) as the Lagrange multiplier to get the Gauss law. So, in
both cases only one component in the $A_{\mu }$ may be eliminated.} \cite%
{ogi3}. It is also well-known \cite{GL} that there is no way to construct a
massless field $A_{\mu }$, which transforms properly as a 4-vector, as a
linear combination of creation and annihilation operators for helicity $\pm
1 $ states.

Under this assumption of not getting too many constraints\footnote{%
The fact that there is a threat of too many supplementary conditions (an
inconsistency) is because we have chosen not to put a Lagrange multiplier
term for the constraint (\ref{con}) into Eq.~(\ref{eqm}). Had we explicitly
introduced such a Lagrange multiplier term, $F(x)(A^{2}-n^{2}M^{2})$, into
the Lagrangian $L$, the equation of motion for the vector field $A_{\mu }$
would have changed, so that the 4-divergence of this equation would now
determine the Lagrange multiplier function $F(x)$ rather than satisfy the
identity (\ref{id}) appearing below.}, we shall now derive gauge invariance.
Since the 4-divergence of the vector field Euler equation (\ref{eqm}) should
be zero if the equations of motion are used, it means that this divergence
must be expressible as a sum over the equations of motion multiplied by
appropriate quantities. This implies that, without using the equations of
motion but still using the constraint (\ref{con}), we have an identity for
the vector and matter (fermion field, for definiteness) fields of the
following type: 
\begin{eqnarray}
\partial _{\mu }\left( \frac{\partial L}{\partial A_{\mu }}-\partial _{\nu }%
\frac{\partial L}{\partial (\partial _{\nu }A_{\mu })}\right) &\equiv
&\left( \frac{\partial L}{\partial A_{\mu }}-\partial _{\nu }\frac{\partial L%
}{\partial (\partial _{\nu }A_{\mu })}\right) (c)A_{\mu }+  \notag \\
&&+\left( \frac{\partial L}{\partial \psi }-\partial _{\nu }\frac{\partial L%
}{\partial (\partial _{\nu }\psi )}\right) (it)\psi +  \label{id} \\
&&+\overline{\psi }(-it)\left( \frac{\partial L}{\partial \overline{\psi }}%
-\partial _{\nu }\frac{\partial L}{\partial (\partial _{\nu }\overline{\psi }%
)}\right) .  \notag
\end{eqnarray}%
Here the coefficients $c$ and $t$ of the Eulerians on the right-hand side
(which vanish by themselves when the equations of motion are fulfilled) are
some dimensionless constants whose particular values are conditioned by the
starting Lagrangian $L(A_{\mu },\psi )$ taken, for simplicity, with
renormalisable coupling constants. This identity (\ref{id}) implies the
invariance of $L$ under the vector and fermion field local transformations
whose infinitesimal form is given by\footnote{%
Actually, one can confirm this proposition by expanding the action with the
transformed Lagrangian density $\int d^{4}xL(A^{\prime },\psi ^{\prime })$
in terms of functional derivatives and then using the identity equation (\ref%
{id}).}%
\begin{equation}
\delta A_{\mu }=\partial _{\mu }\omega +c\omega A_{\mu },\text{ \ \ }\delta
\psi \text{\ }=it\omega \psi  \label{trans}
\end{equation}%
where $\omega (x)$ is an arbitrary function, only being restricted by the
requirement to conform with the nonlinear constraint (\ref{con}).
Conversely, the identity (\ref{id}) in its turn follows from the invariance
of the Lagrangian $L$ under the transformations (\ref{trans}). Both direct
and converse assertions are in fact particular cases of Noether's second
theorem \cite{noeth}. Apart from this invariance, one has now to confirm
that the transformations (\ref{trans}) in fact form an Abelian symmetry
group. Constructing the corresponding Lie bracket operation $(\delta
_{1}\delta _{2}-\delta _{2}\delta _{1})$ for two successive vector field
variations we find that, while the fermion transformation in (\ref{trans})
is an ordinary Abelian local one with zero Lie bracket, for the vector field
transformations there appears a non-zero result 
\begin{equation}
(\delta _{1}\delta _{2}-\delta _{2}\delta _{1})A_{\mu }=c(\omega
_{1}\partial _{\mu }\omega _{2}-\omega _{2}\partial _{\mu }\omega _{1})
\label{SL}
\end{equation}%
unless the coefficient $c=0$. Note also that for non-zero $c$ the variation
of $A_{\mu }$ given by (\ref{SL}) is an essentially arbitrary vector
function. Such a freely varying $A_{\mu }$ is only consistent with a trivial
Lagrangian (i.e. $L=const$). Thus, in order to have a non-trivial
Lagrangian, it is necessary to have $c=0$ and the theory then possesses an
Abelian local symmetry\footnote{%
We will see below (Sec.~3) that non-zero $c$-type coefficients appear in the
non-Abelian internal symmetry case, resulting eventually in a Yang-Mills
gauge invariant theory.}.

Thus we have shown how the choice of a true vacuum conditioned by the SLIV
constraint (\ref{con}) enforces the modification of the Lagrangian $L$, so
as to convert the starting global $U(1)$ charge symmetry into a local one (%
\ref{trans}). Otherwise, the theory would superfluously restrict the number
of degrees of freedom for the vector field and that would be inadmissible.
This SLIV induced local Abelian symmetry (\ref{trans}) now allows the
Lagrangian $L$ to be determined in full. For a minimal theory with
renormalisable coupling constants, it is in fact the conventional QED
Lagrangian which we eventually come to: 
\begin{equation}
L(A_{\mu },\psi )=-\frac{1}{4}F_{\mu \nu }F^{\mu \nu }+\overline{\psi }%
(i\gamma \partial -m)\psi -eA_{\mu }\overline{\psi }\gamma ^{\mu }\psi
\label{lagr1}
\end{equation}%
with the SLIV constraint $A^{2}=n^{2}M^{2}$ imposed on the vector field $%
A_{\mu }$. In the derivation made, we were only allowed to use gauge
transformations consistent with the constraint (2) which now plays the role
of a gauge-fixing term for the resulting gauge invariant theory\footnote{%
As indicated in refs.~\cite{nambu,dir}, the SLIV\ constraint equation for
the corresponding finite gauge function $\omega (x)$, $(A_{\mu
}+\partial_{\mu }\omega)(A^{\mu} + \partial^{\mu}\omega)=n^{2}M^{2}$,
appears to be mathematically equivalent to the classical Hamilton-Jacobi
equation of motion for a charged particle. Thus, this equation should have a
solution for some class of gauge functions $\omega (x)$, inasmuch as there
is a solution to the classical problem.} (\ref{lagr1}). Note that a quartic 
potential $U(A_{\mu})$ of the type discussed in footnote 1 would give 
vanishing contributions on both sides of Eq.~(\ref{id}), when the nonlinear 
constraint (\ref{con}) with the SLIV scale $M^2$ given in the footnote is 
imposed. Furthermore the contribution of such a potential to the Lagrangian 
(\ref{lagr1}) would then reduce to an inessential constant.

One can rewrite the Lagrangian $L(A_{\mu },\psi )$ in terms of the physical
photons now identified as being the SLIV generated vector Goldstone bosons.
For this purpose let us take the following handy parameterization for the
vector potential $A_{\mu }$ in the Lagrangian $L$: 
\begin{equation}
A_{\mu }=a_{\mu }+\frac{n_{\mu }}{n^{2}}(n\cdot A)\qquad (n\cdot A\equiv
n_{\nu }A^{\nu })  \label{par}
\end{equation}%
where $a_{\mu }$ is the pure Goldstonic mode satisfying 
\begin{equation}
\text{\ }n\cdot a=0,\text{\ }\qquad (n\cdot a\equiv n_{\nu }a^{\nu })
\label{sup}
\end{equation}%
while the effective \textquotedblleft Higgs" mode (or the $A_{\mu }$
component in the vacuum direction) is given by the scalar product $n\cdot A$%
. Substituting this parameterization (\ref{par}) into the vector field
constraint (\ref{con}), one comes to the equation for $n\cdot A$: 
\begin{equation}
\text{\ }n\cdot A\text{\ }=(M^{2}-n^{2}a^{2})^{\frac{1}{2}}=M-\frac{%
n^{2}a^{2}}{2M}+O(1/M^{2})  \label{constr1}
\end{equation}%
where $a^{2}=a_{\mu }a^{\mu }$ and taking, for definiteness, the positive
sign for the square root and expanding it in powers of $a^{2}/M^{2}$.
Putting then the parameterization (\ref{par}) with the SLIV constraint (\ref%
{constr1}) into our basic gauge invariant Lagrangian (\ref{lagr1}), one
comes to the truly Goldstonic model for QED. This model might seem
unacceptable since it contains, among other terms, the inappropriately large
Lorentz violating fermion bilinear $eM\overline{\psi }(\gamma \cdot
n/n^{2})\psi $, which appears when the expansion (\ref{constr1}) is applied
to the fermion current interaction term in the Lagrangian $L$ (\ref{lagr1}).
However, due to local invariance of the Lagrangian (\ref{lagr1}), this term
can be gauged away by making an appropriate redefinition of the fermion
field according to 
\begin{equation}
\psi \rightarrow e^{ieM(x\cdot n/n^{2})}\psi   \label{red}
\end{equation}%
through which the $eM\overline{\psi }(\gamma \cdot n/n^{2})\psi $ term is
exactly cancelled by an analogous term stemming from the fermion kinetic
term. So, one eventually arrives at the essentially nonlinear SLIV
Lagrangian for the Goldstonic $a_{\mu }$ field of the type (taken to first
order in $a^{2}/M^{2}$) 
\begin{eqnarray}
L(a_{\mu },\psi ) &=&-\frac{1}{4}f_{\mu \nu }f^{\mu \nu }-\frac{1}{2}\delta
(n\cdot a)^{2}-\frac{1}{4}f_{\mu \nu }h^{\mu \nu }\frac{n^{2}a^{2}}{M}+
\label{NL} \\
&&+\overline{\psi }(i\gamma \partial +m)\psi -ea_{\mu }\overline{\psi }%
\gamma ^{\mu }\psi +\frac{en^{2}a^{2}}{2M}\overline{\psi }(\gamma \cdot
n)\psi .  \notag
\end{eqnarray}%
We have denoted its field strength tensor by $f_{\mu \nu }=\partial _{\mu
}a_{\nu }-\partial _{\nu }a_{\mu }$, while $h_{\mu \nu }=n^{\mu }\partial
^{\nu }-n^{\nu }\partial ^{\mu }$ is a new SLIV oriented differential tensor
acting on the infinite series in $a^{2}$ coming from the expansion of the
effective \textquotedblleft Higgs" mode (\ref{constr1}), from which we have
only included the first order term $-n^{2}a^{2}/2M$ throughout the
Lagrangian $L(a_{\mu },\psi )$. We have also explicitly introduced the
orthogonality condition $n\cdot a=0$ into the Lagrangian through the second
term, which can be treated as the gauge fixing term (taking the limit $%
\delta \rightarrow \infty $). Furthermore we have retained the notation $%
\psi $ for the redefined fermion field.

This nonlinear QED model was first studied on its own by Nambu long ago \cite%
{nambu}. As one can see, the model contains the massless vector Goldstone
boson modes (keeping the massive ``Higgs" mode frozen), and in the limit $%
M\rightarrow \infty $ is indistinguishable from conventional QED taken in
the general axial (temporal or pure axial) gauge. So, for this part of the
Lagrangian $L(a_{\mu},\psi )$ given by the zero-order terms in $1/M$, the
spontaneous Lorentz violation simply corresponds to a non-covariant gauge
choice in an otherwise gauge invariant (and Lorentz invariant) theory.
Remarkably, also all the other (first and higher order in $1/M$) terms in $%
L(a_{\mu},\psi )$ \ (\ref{NL}), though being by themselves Lorentz and $CPT$
violating ones, appear not to cause physical SLIV effects due to strict
cancellations in the physical processes involved. So, the non-linear
constraint (\ref{con}) applied to the standard QED Lagrangian (\ref{lagr1})
appears in fact to be a possible gauge choice, while the $S$-matrix remains
unaltered under such a gauge convention. This conclusion was first confirmed
at the tree level \cite{nambu} and recently extended to the one-loop
approximation \cite{ac}. All the one-loop contributions to the
photon-photon, photon-fermion and fermion-fermion interactions violating
Lorentz invariance were shown to be exactly cancelled with each other, in
the manner observed earlier for the simplest tree-order diagrams. This
suggests that the vector field constraint $A^{2}=n^{2}M^{2}$, having been
treated as a nonlinear gauge choice at the tree (classical) level, remains
as just a gauge condition when quantum effects are taken into account as
well.

To resume let us recall the steps made in the derivation above. We started
with the most general Lorentz invariant Lagrangian $L(A_{\mu},\psi )$,
proposing only a global internal $U(1)$ symmetry for the charged matter
fields involved. The requirement for the vector field equations of motion to
be compatible with the true vacuum chosen by the SLIV (\ref{con}) led us to
the necessity for the identity (\ref{id}) to be satisfied by the Lagrangian $%
L$. According to Noether's second theorem \cite{noeth}, this identity
implies the invariance of the Lagrangian $L$ under the $U(1)$ charge gauge
transformations of all the interacting fields. And, finally, this local
symmetry allows us to completely establish the underlying theory, which
appears to be standard QED (\ref{lagr1}) taken in the nonlinear gauge (\ref%
{con}) or the nonlinear $\sigma $ model-type QED in a general axial gauge -
both preserving physical Lorentz invariance.

\section{Non-Abelian theory}

Now we extend our discussion to the non-Abelian global internal symmetry
case for a general Lorentz invariant Lagrangian $\mathcal{L}(\boldsymbol{A}%
_{\mu },\boldsymbol{\psi })$ for the vector and matter fields involved. This
symmetry is given by a general group $G$ with $D$ generators $t^{\alpha }$ 
\begin{equation}
\lbrack t_{\alpha },t_{\beta }]=ic_{\alpha \beta \gamma }t_{\gamma },\text{
\ }Tr(t_{\alpha }t_{\beta })=\delta _{\alpha \beta }\text{ \ \ }(\alpha
,\beta ,\gamma =0,1,...,D-1)  \label{com}
\end{equation}%
where $c_{\alpha \beta \gamma }$ are the structure constants of $G$. The
corresponding vector fields, which transform according to the adjoint
representation of $G$, are given in the matrix form $\boldsymbol{A}_{\mu }=%
\boldsymbol{A}_{\mu }^{\alpha }t_{\alpha }$. The matter fields (fermions or
scalars) are, for definiteness, taken in the fundamental representation
column $\boldsymbol{\psi }^{\sigma }$ ($\sigma =0,1,...,d-1$) of $G$. Let us
again, as in the above Abelian case, subject the vector field multiplet $%
\boldsymbol{A}_{\mu }^{\alpha }(x)$ to a SLIV constraint of the form 
\begin{equation}
Tr(\boldsymbol{A}_{\mu }\boldsymbol{A}^{\mu })=\boldsymbol{n}^{2}M^{2},\text{
\ \ }\boldsymbol{n}^{2}\equiv \boldsymbol{n}_{\mu }^{\alpha }\boldsymbol{n}%
^{\mu ,\alpha }=\pm 1,  \label{CON}
\end{equation}%
that presumably chooses the true vacuum in a theory. Here, as usual, we sum
over repeated indices. This covariant constraint is not only the simplest
one, but the only possible SLIV condition which could be written for the
vector field multiplet $\boldsymbol{A}_{\mu }^{\alpha }$ and not be
superfluously restricted (see discussion below).

Although we only propose the $SO(1,3)\times G$ invariance of the Lagrangian $%
\mathcal{L}(\boldsymbol{A}_{\mu},\boldsymbol{\psi })$, the chosen SLIV
constraint (\ref{CON}) in fact possesses a much higher accidental symmetry $%
SO(D,3D)$ determined by the dimensionality $D$ of the $G$ adjoint
representation to which the vector fields $\boldsymbol{A}_{\mu }^{\alpha }$
belong\footnote{%
Actually, in the same way as in the Abelian case$^{1}$, such a SLIV
constraint (\ref{CON}) might be related to the minimisation of some $%
SO(D,3D) $ invariant vector field potential $\mathcal{U}(\boldsymbol{A}_{\mu
})=-\frac{m_{A}^{2}}{2}\,Tr(\boldsymbol{A}_{\mu }\boldsymbol{A}^{\mu })+%
\frac{\lambda _{A}}{4}[Tr(\boldsymbol{A}_{\mu }\boldsymbol{A}^{\mu })]^{2}$
followed by taking the limit $m_{A}^{2}\rightarrow \infty ,$ $\lambda
_{A}\rightarrow \infty $ (while keeping the ratio $m_{A}^{2}/\lambda _{A}$
finite). Notably, the inclusion into this potential of another possible,
while less symmetrical, four-linear self-interaction term of the type $%
(\lambda _{A}^{\prime }/4)Tr(\boldsymbol{A}_{\mu }\boldsymbol{A}^{\mu }%
\boldsymbol{A}_{\nu }\boldsymbol{A}^{\nu })$ would lead, as one can easily
confirm, to an unacceptably large number ($4D$) of vector field constraints
at the potential minimum.}. This symmetry is indeed spontaneously broken at
a scale $M$ 
\begin{equation}
<\boldsymbol{A}_{\mu }^{\alpha }(x)>\text{ }=\boldsymbol{n}_{\mu }^{\alpha }M
\label{vev}
\end{equation}%
with the vacuum direction given now by the `unit' rectangular matrix $%
\boldsymbol{n}_{\mu }^{\alpha }$ describing simultaneously both of the
generalized SLIV cases, time-like ($SO(D,3D)$ $\rightarrow SO(D-1,3D)$) or
space-like ($SO(D,3D)$ $\rightarrow SO(D,3D-1)$) respectively, depending on
the sign of $\boldsymbol{n}^{2}\equiv \boldsymbol{n}_{\mu }^{\alpha }%
\boldsymbol{n}^{\mu ,\alpha }=\pm 1$. This matrix has in fact only one
non-zero element for both cases, subject to the appropriate $SO(D,3D)$
rotation. They are, specifically, $\boldsymbol{n}_{0}^{0}$ or $\boldsymbol{n}%
_{3}^{0}$ provided that the vacuum expectation value (\ref{vev}) is
developed along the $\alpha =0$ direction in the internal space and along
the $\mu =0$ or $\mu =3$ direction respectively in the ordinary
four-dimensional one. As we shall soon see, in response to each of these two
breakings, side by side with one true vector Goldstone boson corresponding
to the spontaneous violation of the actual $SO(1,3)\otimes G$ symmetry of
the Lagrangian $\mathcal{L}$, $D-1$ vector pseudo-Goldstone bosons (PGB)
related to a breaking of the accidental $SO(D,3D)$ symmetry of the
constraint (\ref{CON}) per se are also produced\footnote{%
Note that in total there appear $4D-1$ pseudo-Goldstone modes, complying
with the number of broken generators of $SO(D,3D)$, both for time-like and
space-like SLIV. From these $4D-1$ pseudo-Goldstone modes, $3D$ modes
correspond to the $D$ three component vector states as will be shown below,
while the remaining $D-1$ modes are scalar states which will be excluded
from the theory. In fact $D-r$ actual scalar Goldstone bosons (where $r$ is
the rank of the group $G$), arising from the spontaneous violation of $G$,
are contained among these excluded scalar states.}. Remarkably, in contrast
to the familiar scalar PGB case \cite{GL}, the vector PGBs remain strictly
massless being protected by the simultaneously generated non-Abelian gauge
invariance. Together with the above true vector Goldstone boson, they just
complete the whole gauge field multiplet of the internal symmetry group $G$.

Let us now turn to the possible supplementary conditions which can be
imposed on the vector fields in a general Lagrangian $\mathcal{L}(%
\boldsymbol{A}_{\mu },\boldsymbol{\psi })$, in order to finally establish
its form. While generally $D$ supplementary conditions may be imposed on the
vector field multiplet $\boldsymbol{A}_{\mu }^{\alpha }$, one of them in the
case considered is in fact the SLIV constraint (\ref{CON}). One might think
that the other conditions would appear by taking 4-divergences of the
equations of motion 
\begin{equation}
\frac{\partial \mathcal{L}}{\partial \boldsymbol{A}_{\mu }^{\alpha }}%
-\partial _{\nu }\frac{\partial \mathcal{L}}{\partial (\partial _{\nu }%
\boldsymbol{A}_{\mu }^{\alpha })}=0,  \label{eqmI}
\end{equation}%
which are determined by a variation of the Lagrangian $\mathcal{L}$. The
point is, however, that due to the $G$ symmetry this operation would lead,
on equal terms, to $D$ independent conditions thus giving in total, together
with the basic SLIV constraint (\ref{CON}), $D+1$ constraints for the vector
field multiplet $\boldsymbol{A}_{\mu }^{\alpha }$ which is inadmissible.
Therefore, as in the above Abelian case, the 4-divergences of the Euler
equations (\ref{eqmI}) should not produce supplementary conditions at all
once the SLIV occurs. This means again that such 4-divergences should be
arranged to vanish (though still keeping the global $G$ symmetry) either
identically or as a result of the equations of motion for vector and matter
fields (fermion fields for definiteness) thus implying that, in the absence
of these equations, there must hold a general identity of the type 
\begin{eqnarray}
\partial _{\mu }\left( \frac{\partial \mathcal{L}}{\partial \boldsymbol{A}%
_{\mu }^{\alpha }}-\partial _{\nu }\frac{\partial \mathcal{L}}{\partial
(\partial _{\nu }\boldsymbol{A}_{\mu }^{\alpha })}\right) &\equiv &\left( 
\frac{\partial \mathcal{L}}{\partial \boldsymbol{A}_{\mu }^{\beta }}%
-\partial _{\nu }\frac{\partial \mathcal{L}}{\partial (\partial _{\nu }%
\boldsymbol{A}_{\mu }^{\beta })}\right) C_{\alpha \beta \gamma }\boldsymbol{A%
}_{\mu }^{\gamma }+  \notag \\
&&+\left( \frac{\partial \mathcal{L}}{\partial \boldsymbol{\psi }}-\partial
_{\nu }\frac{\partial \mathcal{L}}{\partial (\partial _{\nu }\boldsymbol{%
\psi })}\right) (iT_{\alpha })\boldsymbol{\psi }+  \label{id1} \\
&&+\overline{\boldsymbol{\psi }}(-iT_{\alpha })\left( \frac{\partial 
\mathcal{L}}{\partial \overline{\boldsymbol{\psi }}}-\partial _{\nu }\frac{%
\partial \mathcal{L}}{\partial (\partial _{\nu }\overline{\boldsymbol{\psi }}%
)}\right) .  \notag
\end{eqnarray}

The coefficients $C_{\alpha \beta \gamma }$ and $T_{\alpha }$ of the
Eulerians on the right-hand side of the identity (\ref{id1}) can readily be
identified with the structure constants $c_{\alpha \beta \gamma }$ and
generators $t_{\alpha }$ (\ref{com}) of the group $G$. This follows because
the right hand side of the identity (\ref{id1}) must transform in the same
way as the left hand side, which transforms as the adjoint representation of 
$G$. Note that these coefficients consist of dimensionless constants
corresponding to the starting `minimal' Lagrangian $\mathcal{L}(\boldsymbol{A%
}_{\mu },\boldsymbol{\psi })$ which is taken, for simplicity, with
renormalisable coupling constants. According to Noether's second theorem 
\cite{noeth}, the identity (\ref{id1}) again means the invariance of $%
\mathcal{L}$ under the vector and fermion field local transformations having
the infinitesimal form 
\begin{equation}
\delta \boldsymbol{A}_{\mu }^{\alpha }=\partial _{\mu }\omega ^{\alpha
}+C_{\alpha \beta \gamma }\omega ^{\beta }\boldsymbol{A}_{\mu }^{\gamma },%
\text{ \ \ }\delta \boldsymbol{\psi }\text{\ }=iT_{\alpha }\omega ^{\alpha }%
\boldsymbol{\psi }  \label{trans1}
\end{equation}%
where $\omega ^{\alpha }(x)$ are arbitrary functions only being restricted,
again as in the above Abelian case, by the requirement to conform with the
corresponding nonlinear constraint (\ref{CON}).

Note that the existence of the starting global $G$ symmetry in the theory is
important for our consideration, since without such a symmetry the basic
identity (\ref{id1}) would be written with arbitrary coefficients $C_{\alpha
\beta \gamma }$ and $T_{\alpha }$. Then this basic identity may be required
for only some particular vector field $\boldsymbol{A}_{\mu }^{\alpha _{0}}$
rather than for the entire set $\boldsymbol{A}_{\mu }^{\alpha }$. This would
eventually lead to the previous pure Abelian theory case just for this $%
\boldsymbol{A}_{\mu }^{\alpha _{0}}$ component leaving aside all the other
ones. Just the existence of the starting global symmetry $G$ ensures a
non-Abelian group-theoretical solution for the local transformations (\ref%
{trans1}) in the theory.

So, we have shown that in the non-Abelian internal symmetry case, as well as
in the Abelian case, the imposition of the SLIV constraint (\ref{CON})
converts the starting global symmetry $G$ into the local one $G_{loc}$.
Otherwise, the theory would superfluously restrict the number of degrees of
freedom for the vector field multiplet $\boldsymbol{A}_{\mu }^{\alpha }$,
which would certainly not be allowed. This SLIV induced local non-Abelian
symmetry (\ref{trans1}) now completely determines the Lagrangian $\mathcal{L}
$, following the standard procedure (see, for example, \cite{rabi}). For a
minimal theory with renormalisable coupling constants, this corresponds in
fact to a conventional Yang-Mills type Lagrangian 
\begin{equation}
\mathcal{L}(\boldsymbol{A}_{\mu },\psi )=-\frac{1}{4}\,Tr(\boldsymbol{F}%
_{\mu \nu }\boldsymbol{F}^{\mu \nu })+\overline{\boldsymbol{\psi }}(i\gamma
\partial -m)\boldsymbol{\psi }+g\overline{\boldsymbol{\psi }}\boldsymbol{A}%
_{\mu }\gamma ^{\mu }\boldsymbol{\psi }  \label{nab}
\end{equation}%
(where $\boldsymbol{F}_{\mu \nu }\boldsymbol{~=~}\partial _{\mu }\boldsymbol{%
A}_{\nu }-\partial _{\nu }\boldsymbol{A}_{\mu }-ig[\boldsymbol{A}_{\mu },%
\boldsymbol{A}_{\nu }]$ and $g$ stands for the universal coupling constant
in the theory) with the SLIV constraint (\ref{CON}) imposed. These
constrained gauge fields $\boldsymbol{A}_{\mu }^{\alpha }$ contain, as we
directly confirm below, one true Goldstone and $D-1$ pseudo-Goldstone vector
bosons, corresponding to the spontaneous violation of the accidental $%
SO(D,3D)$ symmetry of the constraint (\ref{CON}).

Actually, as in the above Abelian case, after the explicit use of the
corresponding SLIV constraint (\ref{CON}), which is so far the only
supplementary condition for the vector field multiplet $\boldsymbol{A}_{\mu
}^{\alpha }$, one can identify the pure Goldstone field modes $\boldsymbol{a}%
_{\mu }^{\alpha }$ as follows: 
\begin{equation}
\text{\ \ }\boldsymbol{A}_{\mu }^{\alpha }=\boldsymbol{a}_{\mu }^{\alpha }+%
\frac{\boldsymbol{n}_{\mu }^{\alpha }}{\boldsymbol{n}^{2}}(\boldsymbol{n}%
\cdot \boldsymbol{A}),\text{ \ }\boldsymbol{n}\cdot \boldsymbol{a}\equiv 
\boldsymbol{n}_{\mu }^{\alpha }\boldsymbol{a}^{\mu ,\alpha }\text{\ }=0.
\label{sup'}
\end{equation}%
At the same time an effective \textquotedblleft Higgs" mode (i.e., the $%
\boldsymbol{A}_{\mu }^{\alpha }$ component in the vacuum direction $%
\boldsymbol{n}_{\mu }^{\alpha }$) is given by the product $\boldsymbol{n}%
\cdot \boldsymbol{A}\equiv \boldsymbol{n}_{\mu }^{\alpha }\boldsymbol{A}%
^{\mu ,\alpha }$ determined by the SLIV constraint 
\begin{equation}
\text{\ }\boldsymbol{n}\cdot \boldsymbol{A}\text{\ }=\left[ M^{2}-%
\boldsymbol{n}^{2}\boldsymbol{a}^{2}\right] ^{\frac{1}{2}}=M-\frac{%
\boldsymbol{n}^{2}\boldsymbol{a}^{2}}{2M}+O(1/M^{2}).  \label{constr''}
\end{equation}%
where $\boldsymbol{a}^{2}=\boldsymbol{a}_{\nu }^{\alpha }\boldsymbol{a}^{\nu
,\alpha }.$ As earlier in the Abelian case, we take the positive sign for
the square root and expand it in powers of $\boldsymbol{a}^{2}/M^{2}$. Note
that, apart from the pure vector fields, the general Goldstonic modes $%
\boldsymbol{a}_{\mu }^{\alpha }$ contain $D-1$ scalar fields, $\boldsymbol{a}%
_{0}^{\alpha ^{\prime }}$ or $\boldsymbol{a}_{3}^{\alpha ^{\prime }}$ ($%
\alpha ^{\prime }=1...D-1$), for the time-like ($\boldsymbol{n}_{\mu
}^{\alpha }=n_{0}^{0}g_{\mu 0}\delta ^{\alpha 0}$) or space-like ($%
\boldsymbol{n}_{\mu }^{\alpha }=n_{3}^{0}g_{\mu 3}\delta ^{\alpha 0}$) SLIV
respectively. They can be eliminated from the theory if one imposes
appropriate supplementary conditions on the $\boldsymbol{a}_{\mu }^{\alpha
}$ fields which are still free of constraints. Using their overall orthogonality
(\ref{sup'}) to the physical vacuum direction $\boldsymbol{n}_{\mu }^{\alpha
}$, one can formulate these supplementary conditions in terms of a general
axial gauge for the entire $\boldsymbol{a}_{\mu }^{\alpha }$ multiplet 
\begin{equation}
n\cdot \boldsymbol{a}^{\alpha }\equiv n_{\mu }\boldsymbol{a}^{\mu ,\alpha
}=0,\text{ \ }\alpha =0,1,...D-1.  \label{sup''}
\end{equation}%
Here $n_{\mu }$ is the unit Lorentz vector, analogous to that introduced in
the Abelian case, which is now oriented in Minkowskian space-time so as to
be parallel to the vacuum matrix\footnote{%
For such a choice the simple identity $\boldsymbol{n}_{\mu }^{\alpha }\equiv 
\frac{n\cdot \boldsymbol{n}^{\alpha }}{n^{2}}n_{\mu }$ holds, showing that
the rectangular vacuum matrix $\boldsymbol{n}_{\mu }^{\alpha }$ has the
factorized \textquotedblleft two-vector" form.} $\boldsymbol{n}_{\mu
}^{\alpha }$. As a result, apart from the \textquotedblleft Higgs" mode
excluded earlier by the above orthogonality condition (\ref{sup'}), all the
other scalar fields are also eliminated, and only the pure vector fields, $%
\boldsymbol{a}_{i}^{\alpha }$ ($i=1,2,3$ ) or $\boldsymbol{a}_{\mu ^{\prime
}}^{\alpha }$ ($\mu ^{\prime }=0,1,2$) for time-like or space-like SLIV
respectively, are left in the theory. Clearly, the components $\boldsymbol{a}%
_{i}^{\alpha =0}$ and $\boldsymbol{a}_{\mu ^{\prime }}^{\alpha =0}$
correspond to the Goldstone boson, for each type of SLIV respectively, while
all the others (for $\alpha =1...D-1$) are vector PGBs.

We now show that these Goldstonic vector fields, denoted generally as $%
\boldsymbol{a}_{\mu }^{\alpha }$ but with the supplementary conditions (\ref%
{sup''}) understood, appear truly massless in the SLIV inspired gauge
invariant Lagrangian $\mathcal{L}$ (\ref{nab}) subject to the SLIV
constraint (\ref{CON}). Actually, substituting the parameterization (\ref%
{sup'}) with the SLIV constraint (\ref{constr''}) into the Lagrangian (\ref%
{nab}), one is led to a highly nonlinear Yang-Mills theory in terms of the
pure Goldstonic modes $\boldsymbol{a}_{\mu }^{\alpha }$. However, as in the
above Abelian case, one should first use the local invariance of the
Lagrangian $\mathcal{L}$ to gauge away the apparently large Lorentz
violating terms, which appear in the theory in the form of fermion and
vector field bilinears. As one can readily see, they stem from the expansion
(\ref{constr''}) when it is applied to the couplings $g\overline{\boldsymbol{%
\psi }}\boldsymbol{A}_{\mu }\gamma ^{\mu }\boldsymbol{\psi }$ \ and $-\frac{1%
}{4}g^{2}Tr([\boldsymbol{A}_{\mu }\boldsymbol{,A}_{\nu }]^{2})$ respectively
in the Lagrangian (\ref{nab}). Analogously to the Abelian case, we make the
appropriate redefinitions of the fermion ($\boldsymbol{\psi }$) and vector ($%
\boldsymbol{a}_{\mu }\equiv \boldsymbol{a}_{\mu }^{\alpha }t^{\alpha }$)
field multiplets: 
\begin{equation}
\boldsymbol{\psi }\rightarrow U(\omega )\boldsymbol{\psi }\text{ },\text{ \
\ }\boldsymbol{a}_{\mu }\rightarrow U(\omega )\boldsymbol{a}_{\mu }U(\omega
)^{\dagger },\text{ \ }U(\omega )=e^{igM(x\cdot \boldsymbol{n}^{\alpha }/%
\boldsymbol{n}^{2})\boldsymbol{t}^{\alpha }}.  \label{red1}
\end{equation}%
Since the phase of the transformation matrix $U(\omega )$ is linear in the
space-time coordinate, the following equalities are evidently satisfied: 
\begin{equation}
\partial _{\mu }U(\omega )=igM\boldsymbol{n}_{\mu }U(\omega )=igMU(\omega )%
\boldsymbol{n_{\mu }},\text{ \ \ }\boldsymbol{n_{\mu }}\equiv \boldsymbol{n}%
_{\mu }^{\alpha }t^{\alpha }.
\end{equation}%
One can readily confirm that the above-mentioned Lorentz violating terms are
thereby cancelled with the analogous bilinears stemming from their kinetic
terms. So, the final Lagrangian for the Goldstonic Yang-Mills theory takes
the form (to first order in $(\boldsymbol{a}^{2}/M^{2}$) 
\begin{eqnarray}
\mathcal{L}(\boldsymbol{a}_{\mu }^{\alpha },\boldsymbol{\psi }) &=&-\frac{1}{%
4}Tr(\boldsymbol{f}_{\mu \nu }\boldsymbol{f}^{\mu \nu })-\frac{1}{2}%
\boldsymbol{\delta }(n\cdot \boldsymbol{a}^{\alpha })^{2}+\frac{1}{4}Tr(%
\boldsymbol{f}_{\mu \nu }\boldsymbol{h}^{\mu \nu })\frac{\boldsymbol{n}^{2}%
\boldsymbol{a}^{2}}{M}+  \notag \\
&&+\overline{\boldsymbol{\psi }}(i\gamma \partial -m)\boldsymbol{\psi }+g%
\overline{\boldsymbol{\psi }}\boldsymbol{a}_{\mu }\gamma ^{\mu }\boldsymbol{%
\psi }-\frac{g\boldsymbol{n}^{2}\boldsymbol{a}^{2}}{2M}\overline{\boldsymbol{%
\psi }}(\gamma \cdot \boldsymbol{n})\boldsymbol{\psi }.  \label{nab3}
\end{eqnarray}%
Here the tensor $\boldsymbol{f}_{\mu \nu }$ is, as usual, $\boldsymbol{f}%
_{\mu \nu }\boldsymbol{~=~}\partial _{\mu }\boldsymbol{a}_{\nu }-\partial
_{\nu }\boldsymbol{a}_{\mu }-ig[\boldsymbol{a}_{\mu },\boldsymbol{a}_{\nu }]$%
, while $\boldsymbol{h}_{\mu \nu }$ is a new SLIV oriented tensor of the
type 
\begin{equation*}
\boldsymbol{h}_{\mu \nu }=\boldsymbol{n}_{\mu }\partial _{\nu }-\boldsymbol{n%
}_{\nu }\partial _{\mu }+ig([\boldsymbol{n}_{\mu },\boldsymbol{a}_{\nu }]-[%
\boldsymbol{n}_{\nu },\boldsymbol{a}_{\mu }])
\end{equation*}%
acting on the infinite series in $\boldsymbol{a}^{2}$ coming from the
expansion of the effective \textquotedblleft Higgs" mode (\ref{constr''}),
from which we have only included the first order term $-\boldsymbol{n}^{2}%
\boldsymbol{a}^{2}/2M$ throughout the Lagrangian $\mathcal{L}(\boldsymbol{a}%
_{\mu }^{\alpha }\boldsymbol{,\psi })$. We have explicitly introduced the
(axial) gauge fixing term into the Lagrangian, corresponding to the
supplementary conditions (\ref{sup''}) imposed. We have also retained the
original notations for the fermion and vector fields after the
transformations (\ref{red1}).

The theory we here derived is in essence a generalization of the nonlinear
QED model \cite{nambu} for the non-Abelian case. As one can see, this theory
contains the massless vector Goldstone and pseudo-Goldstone boson multiplet $%
\boldsymbol{a}_{\mu }^{\alpha }$ gauging the starting global symmetry $G$
and, in the limit $M\rightarrow \infty $, is indistinguishable from
conventional Yang-Mills theory taken in a general axial gauge. So, for this
part of the Lagrangian $\mathcal{L}(\boldsymbol{a}_{\mu }^{\alpha }%
\boldsymbol{,\psi })$ given by the zero-order terms in $1/M$, the
spontaneous Lorentz violation again simply corresponds to a non-covariant
gauge choice in an otherwise gauge invariant (and Lorentz invariant) theory.
Furthermore one may expect that, as in the nonlinear QED model \cite{nambu},
all the first and higher order terms in $1/M$ in $\mathcal{L}$\ (\ref{nab3}%
), though being by themselves Lorentz and $CPT$ violating ones, do not cause
physical SLIV effects due to the mutual cancellation of their contributions
to the physical processes involved. Recent tree level calculations \cite{cjm}
related to the Lagrangian $\mathcal{L}(\boldsymbol{a}_{\mu }^{\alpha }%
\boldsymbol{,\psi })$ seem to confirm this proposition. Therefore, the SLIV
constraint (\ref{CON}) applied to a starting general Lagrangian $\mathcal{L}(%
\boldsymbol{A}_{\mu }^{\alpha },\boldsymbol{\psi })$, while generating the
true Goldstonic vector field theory for the non-Abelian charge-carrying
matter, is not likely to manifest itself in a physical Lorentz invariance
violating way.

\section{Conclusion}

The spontaneous Lorentz violation realized through a nonlinear vector field
constraint of the type $A^{2}=M^{2}$ ($M$ is the proposed scale for Lorentz
violation) is shown to generate massless vector Goldstone bosons gauging the
starting global internal symmetries involved, both in the Abelian and the
non-Abelian symmetry case. The gauge invariance, as we have seen, directly
follows from a general variation principle and Noether's second theorem \cite%
{noeth}, as a necessary condition for these bosons not to be superfluously
restricted in degrees of freedom\ once the true vacuum in a theory is chosen
by the SLIV constraint. It should be stressed that we can of course only
achieve this derivation of gauge invariance by allowing all the coupling
constants in the Lagrangian density to be determined from the requirement of
avoiding any extra restriction imposed on the vector field(s) in addition to
the SLIV constraint. Actually, this derivation excludes \textquotedblleft
wrong\textquotedblright\ couplings in the vector field Lagrangian, which
would otherwise distort the final Lorentz symmetry broken phase with
unphysical extra states including ghost-like ones. Note that this procedure
might, in some sense, be inspired by string theory where the coupling
constants are just vacuum expectation values of the dilaton and moduli
fields \cite{string}. So, the adjustment of coupling constants in the
Lagrangian would mean, in essence, a certain choice for the vacuum
configurations of these fields, which are thus correlated with the SLIV.
Another important point for this gauge symmetry derivation is that we
followed our philosophy of imposing the SLIV constraints, (\ref{con}) and (%
\ref{CON}) respectively, without adding a Lagrange multiplier term, as one
might have imagined should come with these constraints. Had we done so the
equations of motion would have changed and the Lagrange multiplier might
have picked up the inconsistency, which we required to be solved in the
Abelian case by Eq.~(\ref{id}) and in the non-Abelian case by Eq.~(\ref{id1}%
).

In the Abelian case a massless vector Goldstone boson appears, which is
naturally associated with the photon. In the non-Abelian case it was shown
that the pure Lorentz violation still generates just one genuine Goldstone
vector boson. However the SLIV constraint (\ref{CON}) manifests a larger
accidental $SO(D,3D)$ symmetry, which is not shared by the Lagrangian $%
\mathcal{L}$. The spontaneous violation of this $SO(D,3D)$ symmetry
generates $D-1$ pseudo-Goldstone vector bosons which, together with the
genuine Goldstone vector boson, complete the whole gauge field multiplet of
the internal symmetry group $G$. Remarkably, these vector bosons all appear
to be strictly massless, as they are protected by the simultaneously
generated non-Abelian gauge invariance. These theories, both Abelian and
non-Abelian, though being essentially nonlinear, appear to be physically
indistinguishable from the conventional QED and Yang-Mills theories due to
their generic, SLIV enforced, gauge invariance. One could actually see that
just this gauge invariance ensures that our theories do not have
unreasonably large (proportional to the SLIV scale $M$ ) Lorentz violation
in the fermion and vector field interaction terms. It appears also to ensure
that all the physical Lorentz violating effects, even those suppressed by
this SLIV scale, are non-observable.

In this connection, the only way for physical Lorentz violation then to
appear would be if the above gauge invariance is somehow broken at very
small distances. One could imagine how such a breaking might occur. Only
gauge invariant theories provide, as we have learned, the needed number of
degrees of freedom for the interacting vector fields once the SLIV occurs.
Note that a superfluous restriction on a vector (or any other) field would
make it impossible to set the required initial conditions in the appropriate
Cauchy problem and, in quantum theory, to choose self-consistent equal-time
commutation relations \cite{ogi3}. One could expect, however, that gravity
could in general hinder the setting of the required initial conditions at
extra-small distances. Eventually this would manifest itself in the
violation of the above gauge invariance in a theory through some high-order
operators stemming from the gravity-influenced area, which could lead to
physical Lorentz violation. We may return to this interesting possibility
elsewhere.

\section*{Acknowledgments}

We would like to thank Rabi Mohapatra for useful discussions and comments.

\end{document}